\documentstyle[12pt,aaspp4]{article}

\begin{document}

\vskip -1cm \hskip 10cm {\bf Fermilab-Pub-99-348-A}
\title{Ultra-High Energy Cosmic  Rays from  Young Neutron Star Winds}
\author{P. Blasi}
\affil{NASA/Fermilab Astrophysics Group, Fermi National Accelerator
Laboratory, Box 500, Batavia, IL 60510-0500}
\author{R. I. Epstein}
\affil{NIS-2, Los Alamos National Laboratory, Los Alamos, NM 87545}
\centerline{and}
\author{A. V. Olinto}
\affil{Department of Astronomy \& Astrophysics \& Enrico Fermi
Institute, University of Chicago, Chicago, IL 60637}


\begin{abstract}  
The long-held notion that the highest-energy cosmic
rays are of distant extragalactic origin is challenged by
observations that events above $\sim 10^{20}$ eV do not exhibit the
expected high-energy cutoff from photopion production off the cosmic
microwave background. We suggest that these  unexpected
ultra-high-energy events are due to iron nuclei accelerated from young
strongly magnetized neutron stars through relativistic MHD winds. We
find that neutron stars whose initial spin periods are shorter than
$\sim 10$ ms and  surface
magnetic fields are in the $10^{12} - 10^{14}$ G range can accelerate iron
cosmic rays to greater than
$\sim 10^{20}$ eV.  These ions can pass through the remnant of the
supernova explosion that produced the neutron star without suffering
significant spallation reactions or energy loss. For plausible models
of the Galactic magnetic field, the trajectories of the iron ions curve
sufficiently to be consistent with the observed, largely isotropic
arrival directions of the highest energy events.
\end{abstract}

Subject headings: acceleration of particles, magnetic fields, MHD, plasmas

\vskip 1cm

The  detection of cosmic rays with energies above
$10^{20}$ eV  has triggered considerable interest on  the  origin
and nature of these particles.  Hundreds of events
with energies above $10^{19}$ eV and about 20 events above
$10^{20}$ eV  have now been observed by a number of experiments such
as HiRes (Kieda et al. 1999), AGASA (Takeda et al. 1998,
1999)\markcite{takeda98,takeda99},  Fly's Eye (Bird et al. 1995,
1993, 1994)\markcite{bird}, Haverah Park (Lawrence, Reid \& Watson
1991)\markcite{L91}, Yakutsk (1996), and Volcano Ranch (1963). Most
unexpected is the large flux of events observed above $5 \times
10^{19}$ eV (Takeda et al. 1998)
\markcite{takeda98} with no sign of
the Greisen-Zatsepin-Kuzmin (GZK) cutoff (Greisen 1966;
Zatsepin \& Kuzmin 1966)\markcite{G66,ZK66}. The
cutoff should be present if these ultra-high energy particles are
protons produced by sources distributed homogeneously throughout the
universe. Cosmic ray protons of energy above $5 \times 10^{19}$ eV
lose their energy to photopion production off the cosmic microwave
background and cannot originate further than about $50\,$Mpc away
from us. Alternatively, if ultra-high-energy cosmic rays (UHECRs)
are protons from sources closer than $50\,$Mpc, the arrival
direction of the events should point toward their source. The
present data shows a mostly  isotropic distribution and no sign of
the local distribution of galaxies or of the Galactic disk
 above  $10^{19}$ eV   (Takeda et al. 1999)\markcite{takeda99}.
In sum, the origin of these particles with
energies tens of millions of times greater than any produced in
terrestrial particle accelerators, remains a mystery.

In addition to the difficulties with locating plausible sources of
UHECRs in our nearby universe, there are great difficulties with
finding plausible accelerators for such extremely energetic particles.
Acceleration of cosmic rays in astrophysical plasmas occurs
when the energy of large-scale macroscopic motion, such as shocks and
turbulent flows, is transferred to individual particles. The
maximum   possible energy, $E_{\rm max}$, is estimated by requiring
that the gyro-radius of the particle be contained in the
acceleration region (Hillas 1994)\markcite{hillas1}
and that the acceleration time be
smaller than the time for energy losses. The former condition
 relates $E_{\rm max}$ to  the strength of the
magnetic field, $B$, and the size of the acceleration region, $L$,
such that  $E_{\rm max}
\sim  \, Ze \, B \, L$, where   $Ze$ is
the charge of the particle. For instance, for $E_{\rm max}
\sim10^{20}$ eV and $Z  \sim 1$, the  known astrophysical
sources with reasonable  $B L $ products are neutron stars ($B \sim
10^{12}$ G  and $L \sim 10$ km), active galactic nuclei ($B \sim
10^{4}$  G and
$L \sim 10$ AU), radio galaxies ($B \sim 10^{-5}$ G and $L \sim
10$ kpc), and clusters of galaxies ($B \sim 10^{-6}$ G and $L
\sim 100$ kpc)
(Hillas 1984; Berezinsky et al. 1990)\markcite{hillas1,berez}.
However, energy losses usually prevent acceleration to
$E_{\rm max}$, and no effective mechanism for UHECR acceleration has
been shown for any of these objects
(Blandford 1999; Bhattacharjee \& Sigl 1998; Venkatesan,  Miller \&
Olinto 1997)\markcite{bland,BS98,VMO}.
Here we show that the early evolution of young magnetized neutron stars
in our Galaxy may be responsible for the  flux of cosmic
rays beyond the GZK cutoff. A preliminary study of this
idea can be found in (Olinto, Epstein \& Blasi 1999)\markcite{OEB}.

Neutron stars have been previously
suggested as  possible sources of UHECRs, starting with an early attempt
by Gunn \& Ostriker (1969)\markcite{og69} to the more recent proposal of
Bell (1992)\markcite{bell}.  Thus far, these attempts have failed at
either reaching the highest energies or reproducing the spectrum or the
apparent isotropy of the arrival directions of UHECRs. In the following
we describe our alternative.  We propose that young neutron stars
may accelerate heavy  nuclei to the highest observed energies by
transferring their rotational energy to particle kinetic energy via a
relativistic MHD wind.

Some neutron stars may  begin their life  rotating rapidly ($\Omega
\sim 3000 {\rm ~rad~s}^{-1}$) and with large surface
magnetic fields ($B_S \gtrsim 10^{13}$ G).  The dipole component of
the field decreases as the cube of the distance from the star's
surface
$B(r) = B_S (R_S / r)^3$, where the radius of the star is    $R_S
\simeq 10^6$ cm. As the distance from the star increases, the dipole
field structure cannot be causally  maintained,  and beyond the {\it
light cylinder} radius,
$R_{lc} = c /\Omega$, the field  is mostly  azimuthal, with field lines
spiraling outwards (Michel 1991)\markcite{M91}.
 For   young, rapidly rotating neutron stars,
the light cylinder   is  about ten times the stellar radius,
$R_{lc} = 10^7
\Omega_{3k}^{-1} $  cm,  where $\Omega_{3k} \equiv \Omega / 3000 \,
{\rm \ rad \ s}^{-1}$.

The surface of young neutron stars is composed of iron peak elements
formed during the supernova event. Iron ions can be stripped off the
hot surface of a young neutron star due to strong electric fields
and be present throughout  much of the magnetosphere
(Ruderman \& Sutherland 1975, Arons  \& Scharlemann
1979)\markcite{RS75}. Inside the light cylinder, the magnetosphere
corotates with the star and the iron density has the Goldreich-Julian
value: $n_{GJ}(r) ={B(r) \Omega /( 4 \pi Z e c} )$,
where  $c$ is the speed of light
(Goldreich \& Julian 1969)\markcite{GJ69}. In this estimate, and what
follows, we do not include the trigonometric factors related to the
relative orientation of the magnetic  and rotational axes.

The exact fate of the plasma outside the light cylinder is still a
subject of  debate (Gallant  \& Arons, 1994;  Begelman \& Li,
1994; Chiueh,  Li, \& Begelman, 1998; Melatos \& Melrose,
1996)\markcite{GA94,BL94,CLB98,MM96}. Observations of
the Crab Nebula indicate that most of the rotational energy emitted by
the Crab pulsar is converted into the kinetic energy of particles in a
relativistic wind  (Kennel \& Coroniti 1994; Begelman 1998; Emmering \&
Chevalier 1987)\markcite{KC84,BL98,EC87}.
This conversion may be due to properties of the MHD flow, related to
magnetic reconnection (Coroniti
1990)\markcite{C90}, or a more gradual end of the MHD  limit (Melatos \&
Melrose, 1996)\markcite{MM96}.   Some analytical and numerical
studies show the development of kinetically dominated relativistic
winds (see e.g., Begelman and Li, 1994)\markcite{BL94},  but at present
the theoretical understanding of the wind dynamics is far from
complete.

The basic idea of accelerating plasmas by  the Poynting flux
 was proposed by Weber and Davis (1967)\markcite{wd67} (then called
{\it magnetic slingshot}).  Later,  Michel(1969)\markcite{michel69}
showed that for a perfectly spherical flow the complete conversion
of the magnetic energy into kinetic energy of the flow could
not be   achieved.   However, Begelman and Li
(1994)\markcite{BL94} reconsidered the problem and showed that even
small deviations from a spherical flow could imply an efficient
conversion of the magnetic energy into kinetic energy of the wind  
through  the so-called magnetic nozzle effect, provided the magnetic
field lines have the right geometry.

In the present study we assume that, at least for some
neutron stars most of the magnetic
energy in the wind zone is converted into the flow kinetic
energy of the
particles in the wind and that the rest mass density of the
regions of the wind containing iron ions are not
dominated by electron-positron pairs; that is, the electron-positron
density is less than $\sim 10^5$ times that of the iron ions. With
these assumptions,  the magnetic field in the wind zone decreases  as
$B(r) \lesssim B_{lc} R_{lc} / r$. For surface fields of  $B_S \equiv
10^{13} \, B_{13}$ G, the field at the light cylinder is
$B_{lc} =10^{10}  \,  B_{13} \Omega_{3k}^3$ G. The
maximum energy  of particles that can be contained in the wind near the
light cylinder is
\begin{equation} E_{max} = {Z e B_{lc} R_{lc} \over c}\simeq 8 \times
10^{20}  \, Z_{26} B_{13} \Omega_{3k}^2 \, {\rm eV} \ ,
\label{eq:Emax}
\end{equation} where $Z_{26} \equiv Z/26$. In the rest frame of the
wind, the plasma is relatively cold while in the star's  rest frame
the plasma moves with Lorentz factors  $ \sim 10^9 - 10^{10}$.

The  typical  energy of the accelerated cosmic rays, $E_{cr}$, can
be estimated by considering the  magnetic energy per ion at the
light cylinder $ E_{cr} \simeq B_{lc}^2/ 8 \pi n_{GJ}$. At the
light cylinder
$ n_{GJ} = 1.7 \times 10^{11} \, {B_{13}  \Omega_{3k}^4 / Z}\, {\rm
cm}^{-3}  $  which  gives
\begin{equation}
 E_{cr} \simeq  4 \times 10^{20}\,  Z_{26} B_{13}
\Omega_{3k}^2 \, {\rm eV} \ ,
\label{eq:Ecr}
\end{equation} similar to $E_{max}$ above
(Gallant \& Arons 1994; Begelman 1994)\markcite{GA94,BL94}.

The spectrum of accelerated UHECRs is determined by the evolution of
the  rotational frequency: As the  star spins down, the energy of
the cosmic ray particles ejected with the wind decreases.
 The total fluence of UHECRs between energy $E$ and $E+dE$  is
\begin{equation} N(E) dE =  \frac{\dot {\cal N}}{\dot \Omega}\,
\frac{d\Omega}{dE}\,  dE \ ,
\label{eq:spec}
\end{equation}  where the particle  luminosity  is
\begin{equation}
\dot {\cal N} = \xi \, n_{GJ} \, \pi R_{lc}^2 c = 6 \times 10^{34} \xi
{B_{13}
\Omega_{3k}^2
\over Z_{26}}   \,  {\rm s}^{-1}
\end{equation} and $\xi<1$  is the efficiency  for accelerating
 particles at the light cylinder. The rotation speed decreases due to
electromagnetic and gravitational radiation
(Lindblom, Owen \& Morsink 1998; Andersson, Kokkotas \& Schutz
1999)\markcite{L98,A99}. For
$B_S \gtrsim 10^{13}$ G,  r-mode gravitational radiation is
likely suppressed (Rezzolla, Lamb \& Shapiro 1999)\markcite{LambSha}
and the spin down
may be dominated  by magnetic dipole radiation  given by:
\begin{equation}
 I \Omega \dot \Omega =  - {B_S^2  R_S^6 \Omega^4 \over 6 c^3}\ .
\label{eq:diprad}
\end{equation} For a moment of inertia $I = 10^{45}$ g cm$^2$, the
time derivative of the spin frequency is,
$ \dot \Omega =    1.7  \times 10^{-5}  B_{13}^2
\Omega_{3k}^3 \, {\rm s}^{-1} $, and  Eq. (\ref{eq:Ecr})  gives
\begin{equation}
\frac{dE}{d\Omega}= {1.7 \times 10^{-3} } \ {E \over \Omega_{3k}}  \ .
\end{equation}
Substituting in Eq. (\ref{eq:spec}), the
particle spectrum from each neutron star is
\begin{equation} N(E) = \xi{ \, 5.5 \times 10^{31} \over B_{13}
E_{20}  Z_{26} }  {\rm GeV}^{-1}    \ ,
\end{equation}
where $E= 10^{20}E_{20}\, {\rm eV}$.

Neutron stars are produced in our Galaxy at a rate
$1/\tau$, where $\tau \equiv 100 ~\tau_2$ yr,  and a fraction $\epsilon$
of them have the required magnetic fields, initial spin rates and
magnetic field geometry to allow efficient conversion of magnetic
energy into kinetic energy of the flow. As discussed below,  UHE 
iron nuclei scatter and   diffuse in the Galactic magnetic field. Taking
the confining volume for these particles to be $V_c$ and the lifetime for
confinement to be $t_c$, the UHECR density is  $n(E) = \epsilon N(E)
t_c/\tau V_c$,  and the flux at the surface of the Earth is $F(E) = n(E)
c/4$.  For a characteristic confinement dimension of $R = 10 ~R_1$  kpc
we can write
$V_c = 4 \pi R^3/3$ and  $t_c = QR/c$, where $Q > 1$ is a measure of
the how well the UHECR are trapped. The predicted UHECR flux at the
Earth is
\begin{equation}
F(E)  = 10^{-24} {   \,
\xi \epsilon Q  \over
\tau_2 R_1^2 B_{13} E_{20}  Z_{26} } \, {\rm GeV}^{-1}  {\rm cm}^{-2}
{\rm
s}^{-1} \  \ .
\label{eq:spec2}
\end{equation}

By comparing with observations, we can estimate the
required efficiency
 factor, $\xi \epsilon$. The AGASA experiment finds that the flux at
$10^{20} {\rm eV}$ at Earth is  $ F(E) = 4 \times 10^{-30}~{\rm
GeV}^{-1}  {\rm cm}^{-2} {\rm s}^{-1}$.   Equating this flux with the
estimate of Eq. (\ref{eq:spec2}), we find that  the efficiency factor
only
needs to be $\xi \epsilon \gtrsim 4 \times 10^{-6} Q^{-1}$.  The
smallness of the required efficiency  suggests that young,
Galactic neutron stars can be the source of UHECRs even if only a small
fraction of stars are born with very rapid  spin frequencies and
high magnetic fields.

The observed  energy spectrum of cosmic rays below the expected GZK cutoff
(i.e., between $\sim 10^8\,$eV and $\lesssim  10^{19}\,$eV) has  a
steep  energy dependence
$N(E)  \propto E^{-\gamma}$,
with $\gamma\approx 2.7$ for $E\lesssim 10^{15}$ eV
and $\gamma\approx 3.1$ for $10^{15}\lesssim E({\rm eV})\lesssim 10^{19}$
(Gaisser 1990)\markcite{gaisser}. The events with energy above
$10^{19.5}\,$eV, however,  show a much flatter spectrum with
$1 \lesssim \gamma \lesssim 2$; the drastic change in slope suggests
the emergence of a new component of cosmic rays  at  ultra-high
 energies.  The predicted spectrum of Eq. (\ref{eq:spec2})
 is very flat, $\gamma =1$,  which agrees with
the lower end of the plausible range of  $\gamma$ observed at
ultra-high energies. Propagation effects can produce an energy
dependence of the  confinement parameter
$Q$ and,  correspondingly, a steepening of the spectrum toward the
middle of the observed range $1 \lesssim \gamma \lesssim 2$.

Even though a young neutron star is usually surrounded by the
remnant of the presupernova star, the accelerated particles can
escape the supernova remnant without significant  degradation as the
envelope expands.
 A requirement for
relativistic winds to  supply  UHECRs
 is that the column density of the envelope becomes transparent to
UHE iron nuclei
before the spinning rate of the neutron star decreases to the level
where the star is unable to emit particles of the necessary energy.  

To estimate the evolution of the column density of the envelope,
consider a supernova that imparts
$E_{SN}= 10^{51} {\cal E}_{51}$ erg  to the stellar envelope of mass
$M_{env}= 10  ~M_1 ~{\rm M}_{\odot}$.  The envelope then disperses with
a velocity
$ v_e \simeq \left({2 E_{SN} / M_{env}}\right)^{1/2} = 3 \times 10^8
\left({{\cal E}_{51}/ M_1}\right)^{1/2} {\rm cm\ s}^{-1}$. The column
density of the envelope surrounding the neutron star is
$\Sigma \simeq {M_{env} / 4 \pi R_{eff}^2}
$ where $R_{eff} = R_0 +v_e t$, where
 $R_0$ is the characteristic radius of the  presupernova  star,
$R_0\lesssim 10^{14} ~{\rm cm}$. We now have
\begin{equation}
\Sigma  \simeq   {M_{env}\over 4 \pi \left [ R_0 +v_e t\right]^2}
   = 1.6 \times 10^{16} { M_1^2 {\cal E}_{51}^{-1}\over t^{2} ( 1 +
t_e/t)^2} \,{\rm g\ cm}^{-2}\ ,
\end{equation}
 where $t$ is in seconds, and
$  t_e = {R_0 / v_e} \lesssim 3 \times 10^5 (M_1/{\cal E}_{51})^{1/2} $
s.
The condition for iron nuclei to traverse the supernova envelope
without significant losses is that
$\Sigma
\lesssim 100 \, $ g cm$^{-2}$.   This ``transparency" occurs at times
$ t > t_{tr} = 1.3\times   10^7 M_1 {\cal E}_{51}^{-1/2}\, {\rm s}\gg
t_e$.

As the envelope is being ejected, the neutron star spin is
slowing due to the magnetic dipole radiation,  Eq.
(\ref{eq:diprad}), so that
\begin{equation}
\Omega_{3k}^{2}(t) = {\Omega_{i3k}^{2} \over  [ 1 + t_8 B_{13}^2
\Omega_{i3k}^2  ]}\ ,
\end{equation} where   $ 3000 ~\Omega_{i3k}$  rad s$^{-1}$ is the
initial spin rate  and  $t_8 = t / 10^8 {\rm s}$. The cosmic ray energy
thus evolves according to
\begin{equation} E_{cr}(t) =   4 \times 10^{20} {\rm eV}\  { Z_{26}
B_{13} \Omega_{i3k}^2
  \over  [ 1 +   t_{8}  B_{13}^2
\Omega_{i3k}^2  ] } \ .
\end{equation}
 The condition that a young neutron star could
produce the UHECRs is that
$E_{cr}$ exceeds the needed energy when the envelope becomes
transparent; i.e.,
$E_{cr}(t_{tr}) > 10^{20} E_{20}$ eV.  This translates into
  the following  condition:
\begin{equation}
 \Omega_{i} > {3000\, {\rm s^{-1}} \over  B_{13}^{1/2} \left [ 4
Z_{26} E_{20}^{-1} -  0.13 M_1 B_{13} {\cal E}_{51}^{-1/2}
\right]^{1/2}}
\ .
\end{equation} From  this equation we obtain the allowed
regions in the $B_S$-$\Omega_i$ plane shown in Figure 1 for $E_{20} =
1$ and 3 and $M_{env}=  5$ and $50 \, {\rm M}_{\odot}$.

For the parameters within the allowed region, the acceleration and
survival of UHE iron nuclei is not significantly affected by the
ambient photon radiation.  The most important source of radiation in
the wind region is the thermal emission from the star's surface. The low
energy non-thermal radiation from the neutron star
is not significant unless it
is $> 10^{4}$ that of the Crab pulsar. In the time needed for the envelope
to become transparent, the surface cools to
$\sim 3\times 10^6$ K (Tsuruta 1998)\markcite{tsuruta}.  For these
temperatures, photodissociation   (see, e.g.,
Protheroe, Bednarek \& Luo 1998\markcite{Proth}) and Compton
drag have minor effects on the energy and composition of the
accelerating iron nuclei. Furthermore, synchrotron losses are
unimportant
because the plasma is essentially cold in the rest frame of the
accelerating plasma.

The  relativistic MHD wind from a rapidly spinning neutron star may
impart more energy to the supernova remnant than the initial explosion.
For initial spin rates $\gtrsim 1000$ rad s$^{-1}$, the rotational
energy is $\gtrsim 10^{51}$ erg, comparable to the kinetic energy of
most supernova remnants.  More rapidly spinning neutron stars
may generate
 highly-energetic supernova events, possibly  similar to SN 1998bw
(Kulkarni 1998)\markcite{Kul}.  In these cases, the  right
boundary of the allowed region in Figure 1 should be enlarged because
the remnant expands more rapidly than assumed above.

The iron  ejected with energies   $\sim 10^{20}$ eV will reach Earth
after being deflected by the Galactic and
halo magnetic  fields (Zirakashvili, Pochepkin, Ptuskin \& Rogovaya
1998)\markcite{Z98}.  The gyroradius of these UHECRs  in the Galactic
field of strength $B_{gal}$ is
\begin{equation} r_B =  {E_{cr}\over Z e B} = {1.4 \over  Z_{26}}
\left({3 \mu {\rm G}
 \over B_{gal}}\right) E_{20} {\rm \ kpc}
\end{equation} which is considerably less than the typical
distance to a young neutron star ($\sim $ 8 kpc).  Therefore,
ultra-high energy iron arriving at the Earth would not point  at the
source.  A  Galactic  iron source is consistent with an approximately
isotropic arrival direction distribution as observed by AGASA for
UHECRs  (Zirakashvili et al. 1998)\markcite{Z98}.
In support of this interpretation, we note that the cosmic ray
component at $10^{18}$ eV is nearly isotropic  with only  a slight
correlation with the Galactic disk and spiral arms (Hayashida et al.
1999)\markcite{H99}. If these cosmic rays are protons of Galactic origin,
 their isotropy  is indicative
of the diffusive effect of the Galactic and halo magnetic fields.  
Since the iron arrival distribution  at $10^{20}$ eV probes  similar
trajectories to protons at a few times $10^{18}$ eV  we expect the
iron  to show a nearly isotropic distribution with a slight
correlation with the Galactic center and disk. This correlation should
become apparent if the number of observed events grows by orders of
magnitude or if events with energies higher than the present highest
energies events are detected. Although some indication of a correlation
with the Galactic center for events above $10^{20}$ eV has been recently
reported (Stanev
\& Hillas 1999)\markcite{SH99}, the small number of observed events
limits the significance of this finding.

In conclusion, we  propose  that  ultra-high-energy cosmic ray
events  originate from iron nuclei accelerated  by young,
strongly magnetic, Galactic neutron stars. Iron  from the surface
of newborn neutron stars are accelerated to ultra-high energies by a
relativistic MHD wind. Neutron stars whose initial spin periods are
shorter than $\sim 4 (B_S/10^{13}{\rm G})^{1/2}$ ms can accelerate
iron nuclei to greater than $10^{20}$ eV.  These ions can pass through
the  radiation field near the neutron star and the remnant of the
supernova explosion that produced the neutron star without suffering
significant deceleration or  spallation reactions.

 The best test of this proposal is a unambiguous composition
(mass/charge) determination and a correlation of arrival directions
for events with energies above $10^{20}$ eV  with the Galactic center
and disk. Both  aspects  will be well tested by future
experiments such as the Auger Project (Cronin 1999)\markcite{cronin}
and OWL-Airwatch (Ormes et al., 1997)\markcite{owl}. In addition,  our
model will be severely constrained if
the indication of a small scale clustering among UHECR events  (Uchihori 
et al. 1999)\markcite{uchi} is confirmed by future experiments
to be due to an isotropically distributed set of
discrete sources.

\vskip 1cm {\bf Acknowledgments}

We are grateful to A. Ferrari, C. Ho, F. K. Lamb and  H. Li for helpful
conversations.  This research was partly supported by NSF through
grant AST 94-20759 at the University of Chicago; by NASA grant NAG 5-7092
at Fermilab, and  by the DOE at Fermilab, at LANL through IGPP, and at
the  University of Chicago through grant DE-FG0291  ER40606.


\newpage

{\bf Figure caption}

Fig. 1\par Parameter space for which acceleration and escape of the
accelerated particles through the ejecta are allowed. The solid
lines  refer to particle energy $E_{cr}=10^{20}$ eV and dashed lines
to $E_{cr}= 3\times 10^{20}$ eV. The curves are plotted for two
values of the envelope mass, $M_{env}=50~M_{solar}$ and
$M_{env}=5~M_{solar}$, as indicated. The horizontal line at spin
period $\sim 0.3$ ms indicates the minimum period (maximum angular
speed) allowed for neutron stars (Haensel, Lasota \& Zdunik 1999)
\markcite{HLZ}.


\end{document}